\documentclass[amsmath,amssymb,preprint,superscriptaddress]{revtex4}

\usepackage{graphicx}
\usepackage{acronym}

\newcommand{\beq}{\begin{equation}}
\newcommand{\eeq}{\end{equation}}
\newcommand{\beqa}{\begin{eqnarray}}
\newcommand{\eeqa}{\end{eqnarray}}

\def\tmc{T_\mathrm{MC}}

\def\N{{\cal N}}

\def\Zb{{\cal Z}_c}
\def\pin{p_\mathrm{in}(R)}
\def\pout{p_\mathrm{out}(R)}
\def\pinalpha{p_\mathrm{in}^{(\alpha)}(R)}
\def\poutalpha{p_\mathrm{out}^{(\alpha)}(R)}

\acrodef{RFOT}{random first-order theory}

\begin{document}

\title{Thermodynamic signature of growing amorphous order in
  glass-forming liquids}

\author{G.~Biroli} 
\affiliation{CEA, DSM, Institut de Physique
  Th{\'e}orique, IPhT, CNRS, MPPU, URA2306, Saclay, F-91191
  Gif-sur-Yvette, France.}
\author{J.-P.~Bouchaud}
\affiliation{Science \& Finance, Capital Fund Management, 6 Bd
  Haussmann, 75009 Paris, France.}
\author{A.~Cavagna}
\affiliation{Centre for Statistical Mechanics and Complexity (SMC),
  CNR-INFM, Via dei Taurini 19, 00185 Roma, Italy.}
\author{T.~S.~Grigera}
\affiliation{Instituto de Investigaciones Fisicoqu{\'\i}micas
  Te{\'o}ricas y Aplicadas (INIFTA -- CCT La Plata) and 
  Departamento de F{\'\i}sica, Facultad de Ciencias Exactas,
  Universidad Nacional de La Plata, c.c. 16, suc. 4, 1900 La Plata,
  Argentina and Consejo Nacional de Investigaciones Cient{\'\i}ficas y
  T{\'e}cnicas, Argentina.}  
\author{P.~Verrocchio}
\affiliation{Dipartimento di Fisica, Universit{\`a} di Trento, via
  Sommarive 14, 38050 Povo, Trento, Italy.}

\begin{abstract}
  Although several theories relate the steep slowdown of glass formers
  to increasing spatial correlations of some sort, standard static
  correlation functions show no evidence for this. We present results
  that reveal for the first time a qualitative \emph{thermodynamic}
  difference between the high temperature and deeply supercooled
  equilibrium glass-forming liquid: the influence of boundary
  conditions propagates into the bulk over larger and larger
  lengthscales upon cooling, and, as this \emph{static} correlation
  length grows, the influence decays nonexponentially. Increasingly
  long-range susceptibility to boundary conditions is expected within
  the \ac{RFOT} of the glass transition, but a quantitative account of
  our numerical results requires a generalization of RFOT where the
  surface tension between states fluctuates.
\end{abstract}

\centerline{\it{Celebrating 50 years of Gibbs-Di Marzio}}

\maketitle

Supercooled liquids show a dramatic slowdown of their dynamics upon
cooling (14 decades increase of viscosity \cite{review:ediger96}
within a narrow temperature range) without any obvious structural or
thermodynamic change \cite{neutronscatt:leheny96}.  The \emph{dynamic}
correlation function of a supercooled liquid furthermore becomes
progressively more nonexponential as the temperature is reduced.
Several theories relate these phenomena to increasing spatial
correlations of some sort \cite{glassthermo:gibbs58,
  mosaic:kirkpatrick89, heterogeneities:garrahan02, review:tarjus05}.
However, \emph{static} correlations have so far been unable to
distinguish qualitatively the high temperature and the deeply
supercooled liquids (though indirect evidences have been inferred from
the specific heat \cite{glassthermo:fernandez06} and the linear
dielectric susceptibility \cite{glassthermo:menon95}).  Inspired by
critical phenomena, it is natural to expect that the slowing down of
the dynamics is related to the vicinity of a thermodynamic phase
transition, where some kind of long-range order would set in
\cite{dynamics:montanari06}.  This is the spirit of different recent
theories \cite{mosaic:kirkpatrick89,review:tarjus05,
  mosaic:bouchaud04, heterogeneities:bouchaud05,glassthermo:coluzzi99,
  review:kivelson97}, but appears at odds with others
\cite{heterogeneities:garrahan02, kinetic:toninelli06}, at least at
first sight.  In particular, the crucial physical mechanism at the
root of \acf{RFOT} \cite{mosaic:kirkpatrick89} is the emergence of
long range amorphous order, whose precise definition and quantitative
characterisation is however far from obvious. Dynamic heterogeneities
\cite{review:ediger00} do show a growing \emph{dynamic} correlation
length accompanying the glass transition, both experimentally
\cite{heterogeneities:berthier05} and numerically \cite[and refs.\
therein]{heterogeneities:toninelli05}.  This is certainly a first
important step, but not sufficient to prune down --- even at a
qualitative level --- different theories of the glass transition. In
particular, it is not clear whether this phenomenon is due to an
underlying static or purely dynamic phase transition.
  
The approach followed here is based on the very definition of a
thermodynamic phase transition, where the effect of boundary
conditions becomes long-ranged. The problem is that for glasses there
are no natural boundary conditions, since these should be as `random'
as the bulk amorphous states that they favor.  A possible solution has
been proposed in \cite{mosaic:bouchaud04}, and further discussed in
\cite{dynamics:montanari06}, in the context of \ac{RFOT}, but the
scope and some conclusions of the \emph{gedanken} experiment proposed
in \cite{mosaic:bouchaud04} are more general
\cite{mosaic:jack05,ising:cammarota07}.  Starting from a given
equilibrium configuration, one freezes the motion of all particles
outside a cavity of radius $R$ and studies the thermodynamics of the
mobile particles with boundary conditions imposed by the frozen ones.
Defining a suitable overlap (inside the cavity) between the original
equilibrated configuration and the configurations equilibrated with
the boundary pinning field, the existence of ``order'' on a scale
$\xi$ implies a large overlap (or `point-to-set' correlation) for $R
\ll \xi$ and a small one for $R \gg \xi$.

Such correlation length arises in \ac{RFOT} \cite{mosaic:bouchaud04},
if one assumes (or finds by approximate calculations
\cite{nucleation:franz05, mosaic:dzero05}) that there are many
amorphous metastable states, ${\cal N} \simeq \exp[R^d\Sigma]$ inside
the cavity ($\Sigma(T)$ is the configurational entropy, or
complexity).  The particles in the cavity can be either in the
original state $\alpha$ in which the outside particles have been
frozen or in any of the other states. The probability to flip to a
different state $\gamma$ is determined by the balance between the
free-energy loss due to mismatch between $\alpha$ and $\gamma$,
$\Upsilon(T) R^{\theta}$, and the gain from the entropic term $T R^d
\Sigma(T)$ ($\Upsilon$ is a generalized surface tension and $\theta\le
d-1$). A crossover between a boundary dominated regime (large overlap)
and an entropy dominated regime (small overlap) is expected for
$R=\xi\propto \left(\Upsilon /T\Sigma\right)^{\frac{1}{d-\theta}}$.
Note that this length diverges at the Kauzmann temperature $T_K$ where
$\Sigma(T_K^+) \to 0$.

This \emph{gedanken} experiment was realized numerically by three of
us in ref.~\onlinecite{self:prl07}. This first study indeed suggested
a growing static length, but this length was very small, and no sharp
transition between high and low overlap was observed, in contrast with
expectations based on RFOT. Is RFOT possibly in contradiction with
numerical experiments?  What is the physical meaning of the
increasingly long range susceptibility to boundary conditions if it is
characterized by such a small length?  In order to answer these
crucial questions we have performed new simulations that measure the
\emph{local} overlap at the center of the cavity. In fact, a major
difficulty in interpreting the results of ref. \onlinecite{self:prl07}
is that the overlap was computed as an average over the whole cavity.
As a result, the overlap drop due to the (putative) change of state of
the cavity is mingled with the decay of the overlap expected from a
trivial weakening of the surface pinning field for larger spheres.
This effect is indeed present even in the single state case
\cite{ising:cammarota07} at all temperatures.  As we shall show,
analyzing the behavior of the local overlap at the center of the
cavity yields sharper results which allow us to answer, at least
partially, the above questions.

\section{Behaviour of the overlap at high and low temperature}

We study a soft-sphere model \cite{soft-spheres:bernu87} that we can
equilibrate below the Mode Coupling transition temperature
$\tmc=$0.226 \cite{soft-spheres:roux89} and for large systems (see
methods). After equilibration, several independent reference
configurations are chosen as starting points for runs with all but $M$
particles frozen.  These mobile particles are confined inside a sphere
of radius $R$ such that the inside density equals the bulk density.
After the confined runs reach equilibrium, the local overlap at the
center $q_c(R)$ is measured. To define $q_c(R)$, we partition the
simulation box in many small cubic boxes of side $\ell$, such that the
probability of finding more than one particle in a single box is
negligible. Let $n_i$ equal the number of particles in box $i$, then
\begin{equation}
  q_c(R) = \frac{1}{\ell^3 N_i} \sum_{i\in v} \langle n_i(t_0) \,
  n_i(t_0+\infty)
  \rangle 
\end{equation}
where the sum runs over all boxes within a small volume $v$ at the
center of the sphere, $N_i$ is the number of boxes, and
$\langle\ldots\rangle$ means thermal average. To minimize statistical
uncertainty without losing the local nature of $q_c(R)$ we choose
$N_i=v/\ell^3=125$. Normalization is such that the overlap of two
identical configurations is $1$ on average, while for totally
uncorrelated configurations $q_c=q_0 \equiv \ell^3=0.062876$.

\begin{figure}
  \centering

  \includegraphics[angle=270,width=\columnwidth]{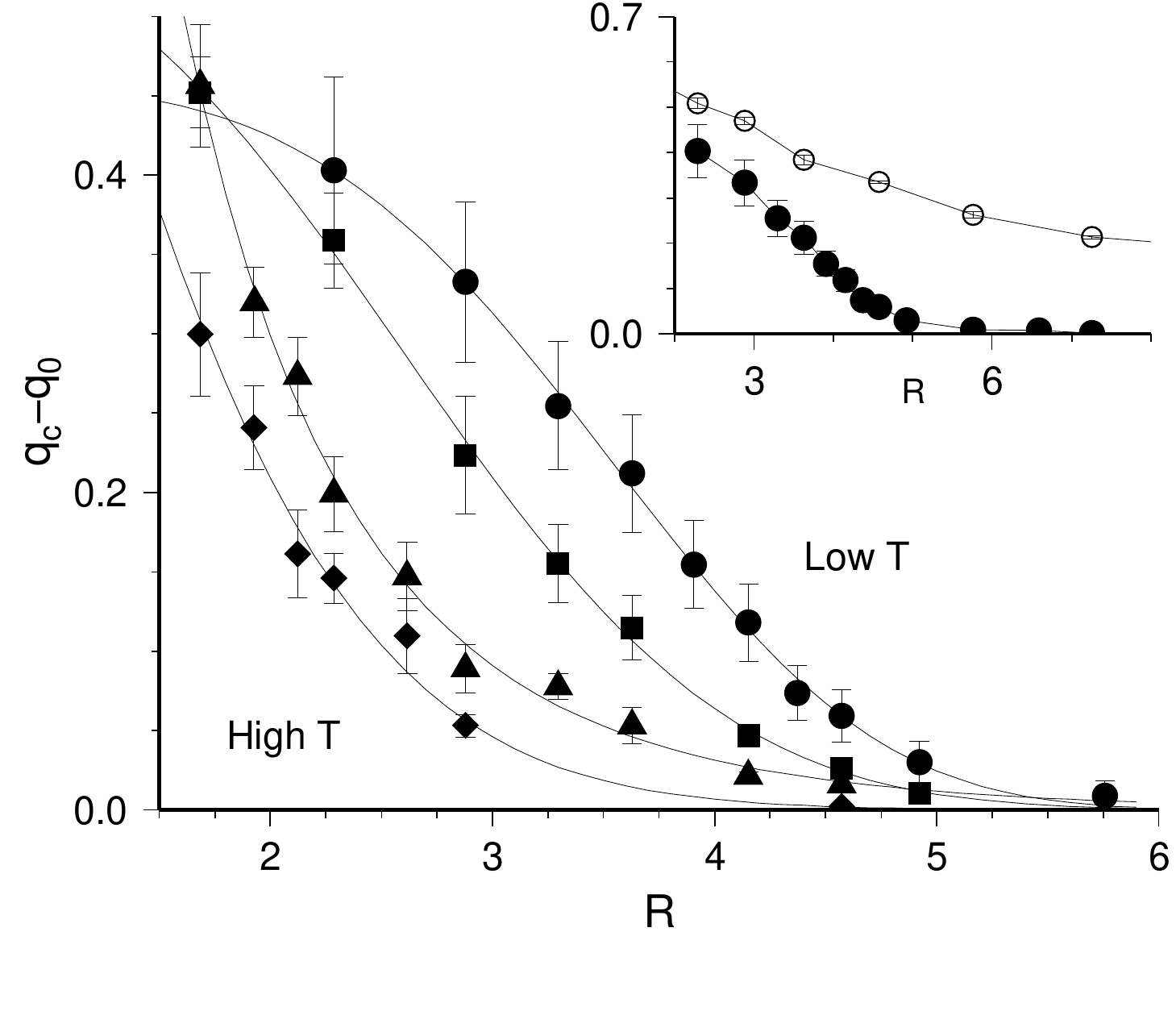}

  \caption{Overlap at the centre of the mobile cavity {\sl vs.\/}
    radius $R$ of the cavity, for temperatures $T=$0.482 (diamonds), 0.350
    (triangles), 0.246 (squares) and 0.203 (circles). Lines are fits
    to eq.~(\ref{pota}). {\bf Inset:} Comparison of $q_c(R)-q_0$ at
    $T=0.203$ (filled circles) with the overlap $Q(R)-q_0$ integrated
    over the whole sphere (open circles, data
    ref.~\onlinecite{self:prl07}).  The local observable $q_c(R)$
    shows a much sharper behaviour.}

\label{fig:alltemps}
\end{figure}

\begin{figure}
  \centering

  \includegraphics[angle=270,width=0.9\columnwidth]{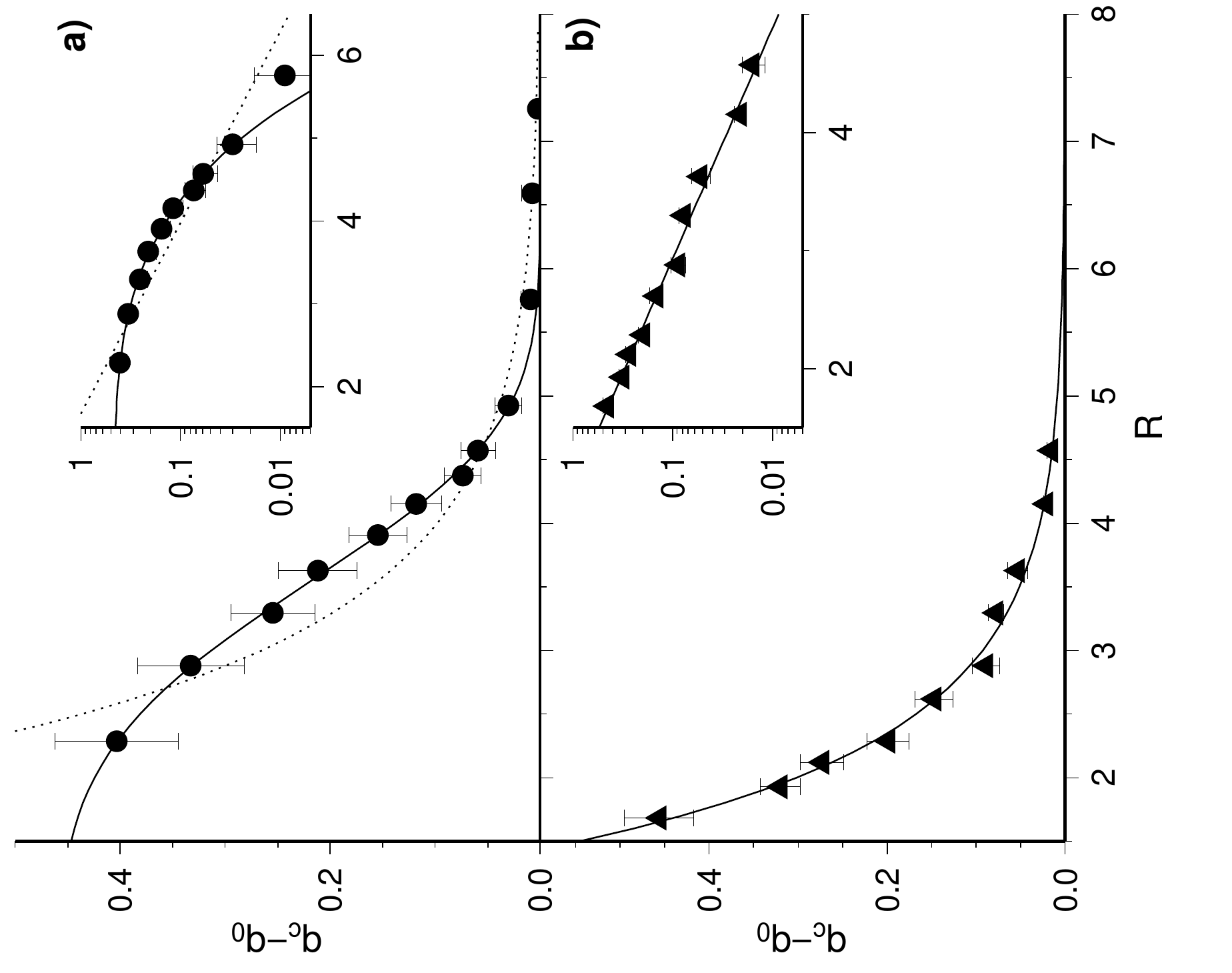}
  
  \caption{{\bf a)} Local overlap at the centre of the mobile cavity
    {\sl vs.\/} $R$ for $T=0.203$ (circles) with fits to a compressed
    exponential (full line) and simple exponential (dotted line). The
    best exponential fit is is rather poor. {\bf Inset:} same data in
    semilog axes, where a simple exponential looks like a straight
    line. {\bf b)} As in panel a) but for $T=0.350$. The exponential
    fit (line) is good at this temperature, as can be seen in the
    inset, which displays the same data in semilog axes. }
  \label{fig:highandlow}
\end{figure}

We show $q_c(R)-q_0$ for several temperatures in
Fig.~\ref{fig:alltemps}.  The decay becomes slower at lower
temperature: the effect of boundary conditions propagates on larger
length-scales. This clearly shows the growth of static order that
expands into the bulk in the deeply supercooled phase. The most
striking result, however, is that at low temperatures the decay is no
longer the simple exponential that prevails in a standard liquid state
\cite{ising:cammarota07}. The relaxation can be fit by a ``compressed
exponential'',
\begin{equation}
 q_c(R)-q_0=\Omega \exp\left[-(R/\xi)^\zeta\right], \qquad \zeta \geq 1,   .  
\label{pota}
\end{equation}
where $\zeta$ increases at low temperatures (see
Fig.~\ref{fig:highandlow}a) above its high temperature liquid value
$\zeta=1$ (see Fig.~\ref{fig:highandlow}b).  Larger values of $\zeta$
means a sharper crossover between large and small overlaps.  The best
fit parameters are given in table~\ref{tab:fitres}. Interestingly, the
value of the length scale $\xi$ found here is significantly larger
than those in \cite{self:prl07}. This is partly due to the fact that
$\zeta > 1$ at low temperatures (see appendix for further discussion).
The overlap $q_c(R)$ therefore appears as a {\it thermodynamic}
quantity able to single out the deeply supercooled liquid in terms of
(a) a large correlation length $\xi$ and (b) an anomalous
nonexponential relaxation, characterized by the exponent $\zeta > 1$.

\begin{table}
  \centering
  \begin{tabular}{c|ccc}
    \hline
    $T$ & $\zeta$ & $\xi$ & $\Omega$ \\
    \hline
    0.482 & 1$^*$ & 0.617(40) & 5.3(1.3) \\
    0.350 & 1$^*$ & 0.845(28) & 3.22(32) \\
    0.246 & 2.79(52) & 3.04(24) & 0.552(61) \\
    0.203 & 4.00(60) & 3.82(12) & 0.46(11) \\
    \hline
  \end{tabular}
  \caption{Best fit parameters as function of temperature. Errors are
    jacknife estimates. (*) For the highest temperatures, the results
    quoted are for a simple exponential fit, which gives a very good
    description of the data with only two parameters. If nonetheless
    those data are fit with the compressed exponential form, an
    exponent $\zeta$ compatible with 1 is obtained, albeit with very
    large (60--80\%) error.}
  \label{tab:fitres}
\end{table}

The growth of $\xi$ strongly suggests that the liquid state should be
thought of as a mosaic of regions (transiently) ordered on a
lengthscale $\xi$.  Furthermore, following suggestions based on
RFOT\cite{mosaic:kirkpatrick89, mosaic:bouchaud04,
  heterogeneities:bouchaud05}, it is natural to conjecture that we are
probing the growth of positional amorphous order. Interestingly, in
this scenario, all finite point static correlations remain
featureless, whereas the \emph{point-to-set} correlations
\cite{dynamics:montanari06, cs:mezard06}, captured by $q_c(R)$, grow.
The simplest version of RFOT predicts at low temperature a sharp
cross-over of $q_c(R)$ for $R=\xi$ \cite{mosaic:franz07}. Our
numerical results show that this is not what happens: the cross-over
is much smoother than expected (see also the recent work
\cite{mosaic:franz07b} on a 1-$d$ Kac model).

\section{RFOT with fluctuating surface tension}

To account for our numerical results, we now propose a natural
generalization of RFOT, based on the idea that the effective interface
tension, $\Upsilon$, is in fact state-dependent. Repeating the
argument of ref.~\onlinecite{mosaic:bouchaud04} with this extra
assumption, the probability that the cavity is found in a state
$\gamma$ different from the outside pinning (frozen) state $\alpha$,
is:
\begin{equation}
  \poutalpha =\frac{\int d\Upsilon e^{R^d \Sigma^* -\beta \Upsilon
      R^{\theta}}p_{\alpha}(\Upsilon|f^*)} {1+\int d\Upsilon e^{R^d
      \Sigma^* -\beta \Upsilon R^{\theta}} p_{\alpha}(\Upsilon|f^*)},
     \qquad \pinalpha = 1-\poutalpha,    \label{eq:pout}
\end{equation}
where $\beta=1/T$, $f^*$ and $\Sigma^*=\Sigma(f^*)$ are
respectively the free energy and configurational entropy of the
equilibrium states, and $p_\alpha(\Upsilon|f)$ is the distribution of
surface tension for state $\alpha$ with other states of free energy
$f$ (see Methods). Since our simulations average over many
realizations of the external state $\alpha$, we in fact compute
$\poutalpha$ averaged over all possible pinning states:
\begin{equation}
  \pout = \sum_\alpha w_\alpha \ \poutalpha =
     \sum_\alpha w_\alpha \ \frac{\int d\Upsilon e^{R^d \Sigma^* -\beta \Upsilon
      R^{\theta}}p_{\alpha}(\Upsilon|f^*)} {1+\int d\Upsilon e^{R^d
      \Sigma^* -\beta \Upsilon R^{\theta}} p_{\alpha}(\Upsilon|f^*)} ,
  \label{eq:pout-noalpha}
\end{equation}
where $w_\alpha$ is the thermodynamic weight of each state. The
leading contribution of the integral over $\Upsilon$ is shown in
Methods to be $e^{-\beta y_\alpha R^\theta }$, where $y_\alpha$ is the
lower edge of $p_{\alpha}(\Upsilon|f)$ and where the exponent $\theta$
is possibly renormalized (this might explain why the effective value
of $\theta$ can be less than the naively expected value $d-1$).
Introducing $P(Y)=\sum_{\alpha}w_{\alpha}\delta(Y-y_{\alpha})$ we find
\begin{equation}
  \pout = \int\!\!dY\,P(Y) \frac{e^{R^d \Sigma^* -\beta Y R^{\theta}}}
                     {1+ e^{R^d \Sigma^* -\beta Y R^{\theta}}}.
\label{norberto}
\end{equation}
The simple mosaic result of ref.~\onlinecite{mosaic:bouchaud04} is
recovered setting $p_{\alpha}(\Upsilon|f) =
\delta(\Upsilon-\Upsilon_0)$ and $P(Y)=\delta(Y-Y_0)$.
Expression~(\ref{norberto}) can be simplified further by noticing that
the approximation $(1+e^{R^d \Sigma^*-\beta Y R^\theta })^{-1} \approx
\Theta(Y-T \Sigma^* R^{d-\theta})$ holds even for rather small values
of $R$ ($\Theta$ is the step function). Hence:
\begin{equation}
  \pout = \int_0^{T \Sigma^* R^{d-\theta}} \!\! P(Y)\,dY. \label{eq:pout-final}
\end{equation}
The overlap at the centre of the cavity is given by $q_c(R) = \pin q_1
+ \pout q_0$, where $q_1$ is the self-overlap of the cavity (which is
not necessarily 1 at finite temperatures). Using
equation~(\ref{eq:pout-final}), we finally obtain
\begin{equation}
q_c(R)-q_0=(q_1-q_0)\int_{T\Sigma^* R^{d-\theta}}^{\infty} dY \, P(Y). 
\label{eq:qc-dist}
\end{equation}
At this point one needs to make some assumption about $P(Y)$.  An
interesting parameterization is:
\begin{equation}
 P(Y) = \frac{\nu}{Y_c}\
\left(\frac{Y}{Y_c}\right)^{\nu-1} \exp\left[-(Y/Y_c)^\nu\right]
= -\frac{d}{dY} \exp\left[-(Y/Y_c)^\nu\right]
\label{eq:alphadistribution}
\end{equation}
($\nu>0$), which describes either a stiff distribution (small values
of $Y$ suppressed) at large $\nu$ or a soft one (small values of $Y$
enhanced) for small $\nu$.  $Y_c$ is the typical value of $\Upsilon$.
The standard mosaic picture corresponds to $\nu \to \infty$.
Equation~(\ref{eq:qc-dist}) now reads, explicitly:
\begin{equation}\label{compressed-exp}
q_c(R)-q_0 = (q_1-q_0) \exp\left[-(R/\xi)^{\nu(d-\theta)}\right]
\end{equation}
where the usual RFOT relation between $\xi$ and $\Sigma$ is recovered:
$\xi=(Y_c/T\Sigma^*)^{1/(d-\theta)}$. As in the Adam-Gibbs treatement
\cite{glassthermo:adam65}, the growth of the length is caused by the
decreasing of the configurational entropy at low temperature.  The
expression (\ref{compressed-exp}) is the compressed exponential form
(\ref{pota}) used above to fit our numerical data. Within this
framework, the thermodynamic anomaly $\zeta=\nu(d-\theta)$ is directly
related (at low temperatures) to the exponent $\nu$ describing the
surface tension distribution.

A non trivial $P(Y)$ could arise for two main reasons. One is strong
pre-asymptotic corrections to standard RFOT: though the surface
tension between two typical states could be self-averaging in the
$R\to\infty$ limit, fluctuations can be state dependent and important
at finite $R$. This is often what happens for interfaces in random
media (see SI). The second possibility is that the effective interface
tension fluctuates even in the large $R$ limit.  If the exponent
$\theta$ turns out to be less than $d-1$, as suggested by
phenomenological arguments in ref.~\onlinecite{mosaic:kirkpatrick89},
then the interface should be a highly fluctuating object as e.g. in
the droplet theory of spin-glasses\cite{spin-glass:fisher86}. We find
that $\zeta$, and therefore $\nu$, increase when $T$ decreases
indicates that the crossover of $q_c(R)$ becomes sharper at larger
sizes (see Fig.~(\ref{fig:alltemps}).  This is compatible with a
finite-size effect scenario.  This behaviour is also expected within
the RFOT scenario that predicts a vanishing surface tension at the
mode-coupling transition $\tmc$, which behaves as a spinodal point.
Coherent amorphous order droplets should therefore be fractal around
$\tmc$ and compact below \cite{mosaic:stevenson06}, which suggests an
increase of the effective value of $\nu$ as $T$ decreases. A first
principle RFOT computation of $q_c(R)$ for the model we simulated
would be very instrumental to clarify this issue.

\section{Conclusions and outlook}

We have unveiled a qualitative difference between the high temperature
and deeply super-cooled equilibrium regimes: the influence of boundary
conditions propagates into the bulk on an increasingly large
lengthscale upon cooling. Furthermore, the growth of this length is
accompanied by a sharpening of the decay.  We have developed a
theoretical framework, based on a generalization of RFOT, that
explains these results as a one-state to multi-state transition
governed by the surface tension distribution. The sharpening of the
decay at low temperature corresponds to more and more regions
developing a large surface tension.  From a more general perspective,
our numerical results strongly support a mosaic picture where the
super-cooled liquid is characterized by a ``hidden'' static order
\footnote{We call it ``hidden'' since, as discussed in the
  introduction, all simple static correlation functions investigated
  until now have never shown direct evidence of a growing length.} on
an increasingly larger scale upon cooling.  Although this is a natural
consequence of RFOT, other theoretical approaches may also account for
these phenomena at least on a qualitative level: in particular the
frustration limited domain theory \cite{review:kivelson97} and,
perhaps surprisingly, some kinetically constrained models (see
\cite{mosaic:jack05} for a discussion of this point). The main
difference is the physical origin of the growing static lengthscale:
within RFOT and at variance with other approaches, it is tightly
linked to the decrease of the configurational entropy. Our work opens
the way to a quantitative study of this issue and, hence, to a
clear-cut test of RFOT as a valid theory of the glass transition. From
a more theoretical point of view, several crucial questions remain
elusive: can a RFOT-like transition exist outside mean-field? How
precisely can amorphous metastable states be defined? Does the notion
of effective surface tension between these states make sense? A
definitive test of the mosaic scenario requires to find a way to
measure \emph{directly} this surface tension and its distribution in
the deeply supercooled phase.  Work in this direction is in progress.

\acknowledgments

We thank C.~Cammarota, L.~A.~Fernandez, G.~Gradenigo, I.~Giardina,
A.~Lef{\`e}vre, V.~Mart{\'\i}{}n-Mayor, A.~Montanari, G.~Parisi,
D.~Reichman, M.~Tarzia, and F.~Zamponi for useful discussions. GB and
JPB are supported by ANR Grant DYNHET. TSG thanks ECT* and
Dipartimento di Fisica, Universit\'a di Trento for hospitality and
partial support and acknowledges partial support from CONICET and
ANPCyT (Argentina) and ICTP (Trieste, Italy).

\section{Methods}

\subsection{Simulation}

We have studied the soft-sphere binary mixture
\cite{soft-spheres:bernu87}, a fragile model glass-former. In addition
to capturing the essential features of fragile glasses, this model
can be thermalized below the Mode Coupling temperature with the the
swap Monte Carlo algorithm of \onlinecite{self:pre01}. Particles are
of unit mass and belong to one of two species $\mu= 1, 2$, present in
equal amounts and interacting via a potential
$$ V = \sum_{i>j}^N v_{ij}(|\mathbf{r}_i-\mathbf{r}_j|) =
       \sum_{i>j}^N \left[ \frac{\sigma_{\mu(i)} + \sigma_{\mu(j)}}
         {|\mathbf{r}_i-\mathbf{r}_j|} \right]^{12},
$$
where the radii $\sigma_\mu$ are fixed by the conditions
$\sigma_2/\sigma_1=1.2$, $(2\sigma_1)^3 + 2(\sigma_1 +\sigma_2)^3 +
(2\sigma_2)^3 = 4\ell_0^3$, and $\ell_0$ is the unit of length. The particle
density is $\rho=N/V=l_0^{-3}$. A smooth long-range cut-off is imposed
setting $v_{ij}(r)=B_{ij}(a-r)^3 + C_{ij}$ for $r>r_c=\sqrt{3}$ and
$v_{ij}(r)=C_{ij}$ for $r>a$, where $a$, $B_{ij}$, and $C_{ij}$ are
fixed by requiring continuity up to the second derivative of
$v_{ij}(r)$. Temperature is measured in units of energy. To obtain the
reference configurations, 4 to 8 replicas of systems with $N=2048$ or
$N=16384$ where equilibrated in a cubic box with periodic boundary
conditions. Then the overlap was computed in systems with frozen
boundaries and $M$ mobile particles, with $M=$20, 30, 40, 50, 100,
150, 200, 300, 400, 800, 1600, 3200 mobile particles ($1.684 \le R \le
9.142$). The results are averaged over 8 to 32 outer states (reference
configurations).  Data were collected for at least 10 relaxation times
(up to $10^6$ Monte Carlo steps), after discarding an initial portion
of at least one relaxations time.  Error bars were obtained from a
jacknife estimate from sample-to-sample fluctuations.

\vskip 1 truecm

\subsection{Overlap}

To obtain equation~(\ref{eq:pout}), write the partition function for
the mobile cavity surrounded by pinning state $\alpha$
\begin{equation}
  \Zb = e^{-\beta R^d f_\alpha} + \sum_{\gamma\neq\alpha} e^{-\beta R^d
    f_\gamma - \beta R^\theta \Upsilon_{\alpha\gamma}},
\end{equation}
so that
\begin{equation}
  \poutalpha = \frac{\sum_{\gamma\neq\alpha} e^{-\beta R^d
    f_\gamma - \beta R^\theta \Upsilon_{\alpha\gamma}}}{\Zb}.
\end{equation}
Introducing $\N_\alpha(f,\Upsilon) = \sum_\gamma \delta(f-f_\gamma)
\delta(\Upsilon - \Upsilon_{\alpha\gamma})$, the sum can be written
\begin{eqnarray}
  \sum_{\gamma\neq\alpha} e^{-\beta R^d f_\beta - \beta R^\theta
    \Upsilon_{\alpha\gamma}} &=& \int\!\!df\!\!\int\!\!d\Upsilon\,
  e^{-\beta R^d f - \beta R^\theta \Upsilon} \N(f,\Upsilon), \\
  & =&
  \int\!\!df\!\!\int\!\!d\Upsilon\, 
  e^{-\beta R^d f - \beta R^\theta \Upsilon + R^d \Sigma(f)}
  p_\alpha(\Upsilon|f),
\end{eqnarray}
where in the last equality we have defined $p_\alpha(\Upsilon|f) =
\N_\alpha(f,\Upsilon) / \N(f)$, and $\N(f)=\exp[\Sigma(f)]$ is the
number of states with free energy $f$. Equation~(\ref{eq:pout})
follows approximating the integral over $f$ with the saddle-point
method, which picks $f=f^*$ as the dominant contribution.
$p_\alpha(\Upsilon|f)$ is the fraction of states $\gamma$ (inside the
cavity) with free energy $f$ and effective interface tension
$\Upsilon$, which we assume to be $R$-independent\footnote{Actually,
  some $R$-dependence would not affect the result. The important
  requirement is that this dependence does not lead to values
  exponentially large or small in $R$}.

Now the integral in equations~(\ref{eq:pout})
and~(\ref{eq:pout-noalpha}) can be simplified using the saddle point
method, which is a very good approximation even for rather small
values of $R$. Because of the exponential term in $R^\theta$, the
integral is dominated by the lowest values of $\Upsilon$ supported by
the distribution $p_{\alpha}(\Upsilon|f)$. There are two possibles
cases: (a) If this function has a left edge, {\sl i.e.\/~} it vanishes
for $\Upsilon<y_{\alpha}$, one finds up to subleading terms: $\int d
\Upsilon\; p_{\alpha}(\Upsilon|f) \ e^{-\beta \Upsilon R^\theta
}\simeq e^{-\beta y_\alpha R^\theta }$. Otherwise: (b) there are
arbitrarily small effective tensions $\Upsilon$. Remarkably, in this
case, depending on the form of $p_{\alpha}(\Upsilon|f)$ at small
$\Upsilon$, one can obtain a renormalization of $\theta$. For instance
in the case $p_{\alpha}(\Upsilon|f)\simeq
\exp(-c_{\alpha}/\Upsilon^a)$, where $a$ is a positive exponent, one
finds $\int d \Upsilon\; p_{\alpha}(\Upsilon|f) \ e^{-\beta \Upsilon
  R^\theta }\simeq e^{-\beta y'_\alpha R^{\theta '} }$, where
$y'_\alpha$ is a constant dependent of the outside state $\alpha$ and
temperature and $\theta'=a\theta/(a+1) < \theta$.

\appendix

\section{Comparison with the results of  ref.~22}

Following the same numeric protocol of this work, in \cite{self:prl07}
the influence of the boundary conditions on the {\em total} overlap
within the sphere was studied. It was observed that the decay of
$q_\mathrm{tot}(R)$ is described sufficiently well in the single state
framework and the standard RFOT scenario was ruled out. In this work
we show however that at low enough temperatures neither the standard
RFOT nor the single state scenario account for the local overlap data
and a generalized RFOT theory is presented which is far more
successful.

One might retrospectively wonder if the data in \cite{self:prl07}
could have been used to discriminate between the one-state and the
generalized RFOT scenarios. The answer is no. Allowing for a $R$
dependence of $q_0$ and $q_1$ of the type suggested in
\cite{self:prl07}:
\begin{equation}
q_\mathrm{0;1}(R) = 3(1-q^*_\mathrm{0;1})\left[ \frac{1}{x} -
  \frac{2}{x^2} + \frac{2 
      \left(1- e^{-x}\right)}{x^3} \right] + q^*_\mathrm{0;1}  
\end{equation} 
with $x\equiv R/\lambda_\mathrm{0;1}$, the generalized RFOT prediction
for $q_\mathrm{tot}(R)$ reads now:
\begin{equation} q_\mathrm{tot}(R) = q_0(R) + (q_1(R) - q_0(R))
\; \exp \left[ -\left( R/\xi\right)^{\zeta} \right] 
\label{newmosaictot}
\end{equation} 

In the figure below we show that the generalized RFOT encoded in
(\ref{newmosaictot}) and the one-state prediction given by the formula
(6) of ref.~\onlinecite{self:prl07} fit the total overlap data at
$T=0.203$ at a comparable level of accuracy. Following Occam's razor
principle ({\em ``entia non sunt multiplicanda praeter
  necessitatem''}), one should choose one-state theory, which is the
one with the smallest number of parameters.

\begin{figure}
  \centering
  \includegraphics[angle=270,width=.9\columnwidth]{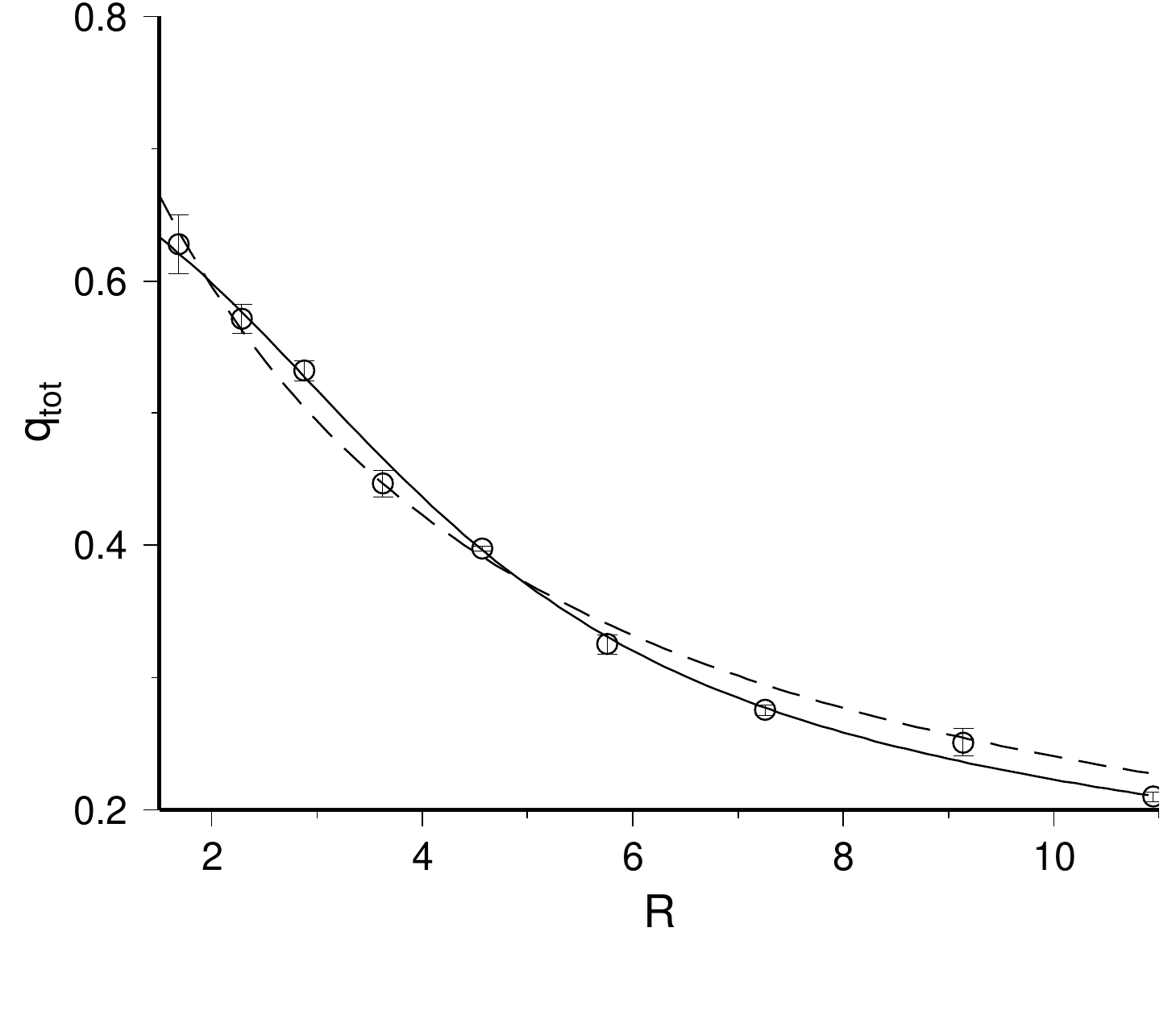}
  \caption{Total overlap of the mobile cavity at $T=0.203$ (data taken
    from ref.~\onlinecite{self:prl07}). The dashed line shows the fit
    with one-state theory while the solid line shows the fit with the
    generalized RFOT prediction (\ref{newmosaictot}).}
\label{figSI}
\end{figure}


\begin{thebibliography}{10}
\expandafter\ifx\csname url\endcsname\relax
  \def\url#1{\texttt{#1}}\fi
\expandafter\ifx\csname urlprefix\endcsname\relax\def\urlprefix{URL }\fi
\providecommand{\bibinfo}[2]{#2}
\providecommand{\eprint}[2][]{\url{#2}}

\bibitem{review:ediger96}
\bibinfo{author}{Ediger, M.~D.}, \bibinfo{author}{Angell, C.~A.} \&
  \bibinfo{author}{Nagel, S.~R.}
\newblock \bibinfo{title}{Supercooled liquids and glasses}.
\newblock \emph{\bibinfo{journal}{J. Phys. Chem.}}
  \textbf{\bibinfo{volume}{100}}, \bibinfo{pages}{13200}
  (\bibinfo{year}{1996}).

\bibitem{neutronscatt:leheny96}
\bibinfo{author}{Leheny, R.~L.} \emph{et~al.}
\newblock \bibinfo{title}{Structural studies of an organic liquid through the
  glass transition}.
\newblock \emph{\bibinfo{journal}{J. Chem. Phys.}}
  \textbf{\bibinfo{volume}{105}}, \bibinfo{pages}{7783--7794}
  (\bibinfo{year}{1996}).

\bibitem{glassthermo:gibbs58}
\bibinfo{author}{Gibbs, J.~H.} \& \bibinfo{author}{DiMarzio, E.~A.}
\newblock \bibinfo{title}{Nature of the glass transition and the glassy state}.
\newblock \emph{\bibinfo{journal}{J. Chem. Phys.}}
  \textbf{\bibinfo{volume}{28}}, \bibinfo{pages}{373--383}
  (\bibinfo{year}{1958}).

\bibitem{mosaic:kirkpatrick89}
\bibinfo{author}{Kirkpatrick, T.~R.}, \bibinfo{author}{Thirumalai, D.} \&
  \bibinfo{author}{Wolynes, P.~G.}
\newblock \bibinfo{title}{Scaling concepts for the dynamics of viscous liquids
  near an ideal glassy state}.
\newblock \emph{\bibinfo{journal}{Phys. Rev. A}} \textbf{\bibinfo{volume}{40}},
  \bibinfo{pages}{1045--1054} (\bibinfo{year}{1989}).

\bibitem{heterogeneities:garrahan02}
\bibinfo{author}{Garrahan, J.~P.} \& \bibinfo{author}{Chandler, D.}
\newblock \bibinfo{title}{Geometrical explanation and scaling of dynamical
  heterogeneities in glass forming systems}.
\newblock \emph{\bibinfo{journal}{Phys. Rev. Lett.}}
  \textbf{\bibinfo{volume}{89}}, \bibinfo{pages}{035704}
  (\bibinfo{year}{2002}).

\bibitem{review:tarjus05}
\bibinfo{author}{Tarjus, G.}, \bibinfo{author}{Kivelson, S.~A.},
  \bibinfo{author}{Nussinov, Z.} \& \bibinfo{author}{Viot, P.}
\newblock \bibinfo{title}{The frustration-based approach of supercooled liquids
  and the glass transition: a review and critical assessment}.
\newblock \emph{\bibinfo{journal}{J. Phys.: Condens. Matter}}
  \textbf{\bibinfo{volume}{17}}, \bibinfo{pages}{R1143} (\bibinfo{year}{2005}).

\bibitem{glassthermo:fernandez06}
\bibinfo{author}{Fern\'andez, L.~A.}, \bibinfo{author}{Mart\'{\i}n-Mayor, V.}
  \& \bibinfo{author}{Verrocchio, P.}
\newblock \bibinfo{title}{Critical behavior of the specific heat in glass
  formers}.
\newblock \emph{\bibinfo{journal}{Phys. Rev. E}} \textbf{\bibinfo{volume}{73}},
  \bibinfo{pages}{020501} (\bibinfo{year}{2006}).

\bibitem{glassthermo:menon95}
\bibinfo{author}{Menon, N.} \& \bibinfo{author}{Nagel, S.~R.}
\newblock \bibinfo{title}{Evidence for a divergent susceptibility at the glass
  transition}.
\newblock \emph{\bibinfo{journal}{Phys. Rev. Lett.}}
  \textbf{\bibinfo{volume}{74}}, \bibinfo{pages}{1230--1233}
  (\bibinfo{year}{1995}).

\bibitem{dynamics:montanari06}
\bibinfo{author}{Montanari, A.} \& \bibinfo{author}{Semerjian, G.}
\newblock \bibinfo{title}{Rigorous inequalities between length and time scales
  in glassy systems}.
\newblock \emph{\bibinfo{journal}{J. Stat. Phys.}}
  \textbf{\bibinfo{volume}{125}}, \bibinfo{pages}{23--54}
  (\bibinfo{year}{2006}).

\bibitem{mosaic:bouchaud04}
\bibinfo{author}{Bouchaud, J.-P.} \& \bibinfo{author}{Biroli, G.}
\newblock \bibinfo{title}{On the
  {A}dam-{G}ibbs-{K}irkpatrick-{T}hirumalai-{W}olynes scenario for the
  viscosity increase in glasses}.
\newblock \emph{\bibinfo{journal}{J. Chem. Phys.}}
  \textbf{\bibinfo{volume}{121}}, \bibinfo{pages}{7347--7354}
  (\bibinfo{year}{2004}).

\bibitem{heterogeneities:bouchaud05}
\bibinfo{author}{Bouchaud, J.-P.} \& \bibinfo{author}{Biroli, G.}
\newblock \bibinfo{title}{Nonlinear susceptibility in glassy systems: A probe
  for cooperative dynamical length scales}.
\newblock \emph{\bibinfo{journal}{Phys. Rev. B}} \textbf{\bibinfo{volume}{72}},
  \bibinfo{pages}{064204} (\bibinfo{year}{2005}).

\bibitem{glassthermo:coluzzi99}
\bibinfo{author}{Coluzzi, B.}, \bibinfo{author}{Mezard, M.},
  \bibinfo{author}{Parisi, G.} \& \bibinfo{author}{Verrocchio, P.}
\newblock \bibinfo{title}{Thermodynamics of binary mixture glasses}.
\newblock \emph{\bibinfo{journal}{J. Chem. Phys.}}
  \textbf{\bibinfo{volume}{111}}, \bibinfo{pages}{9039--9052}
  (\bibinfo{year}{1999}).

\bibitem{review:kivelson97}
\bibinfo{author}{{Kivelson}, D.}, \bibinfo{author}{{Tarjus}, G.} \&
  \bibinfo{author}{{Kivelson}, S.~A.}
\newblock \bibinfo{title}{{A Viewpoint, Model and Theory for Supercooled
  Liquids}}.
\newblock \emph{\bibinfo{journal}{Progr. Theor. Phys. Supp.}}
  \textbf{\bibinfo{volume}{126}}, \bibinfo{pages}{289--299}
  (\bibinfo{year}{1997}).

\bibitem{kinetic:toninelli06}
\bibinfo{author}{Toninelli, C.}, \bibinfo{author}{Biroli, G.} \&
  \bibinfo{author}{Fisher, D.~S.}
\newblock \bibinfo{title}{Jamming percolation and glass transitions in lattice
  models}.
\newblock \emph{\bibinfo{journal}{Phys. Rev. Lett.}}
  \textbf{\bibinfo{volume}{96}}, \bibinfo{pages}{035702}
  (\bibinfo{year}{2006}).

\bibitem{review:ediger00}
\bibinfo{author}{Ediger, M.~D.}
\newblock \bibinfo{title}{Spatially heterogeneous dynamics in supercooled
  liquids}.
\newblock \emph{\bibinfo{journal}{Annu. Rev. Phys. Chem.}}
  \textbf{\bibinfo{volume}{51}}, \bibinfo{pages}{99--128}
  (\bibinfo{year}{2000}).

\bibitem{heterogeneities:berthier05}
\bibinfo{author}{Berthier, L.} \emph{et~al.}
\newblock \bibinfo{title}{Direct experimental evidence of a growing length
  scale accompanying the glass transition}.
\newblock \emph{\bibinfo{journal}{Science}} \textbf{\bibinfo{volume}{310}},
  \bibinfo{pages}{1797--1800} (\bibinfo{year}{2005}).

\bibitem{heterogeneities:toninelli05}
\bibinfo{author}{Toninelli, C.}, \bibinfo{author}{Wyart, M.},
  \bibinfo{author}{Berthier, L.}, \bibinfo{author}{Biroli, G.} \&
  \bibinfo{author}{Bouchaud, J.-P.}
\newblock \bibinfo{title}{Dynamical susceptibility of glass formers:
  Contrasting the predictions of theoretical scenarios}.
\newblock \emph{\bibinfo{journal}{Phys. Rev. E}} \textbf{\bibinfo{volume}{71}},
  \bibinfo{pages}{041505} (\bibinfo{year}{2005}).

\bibitem{mosaic:jack05}
\bibinfo{author}{Jack, R.~L.} \& \bibinfo{author}{Garrahan, J.~P.}
\newblock \bibinfo{title}{Caging and mosaic length scales in plaquette spin
  models of glasses}.
\newblock \emph{\bibinfo{journal}{J. Chem. Phys.}}
  \textbf{\bibinfo{volume}{123}}, \bibinfo{pages}{164508}
  (\bibinfo{year}{2005}).

\bibitem{ising:cammarota07}
\bibinfo{author}{Cammarota, C.} \& \bibinfo{author}{Cavagna, A.}
\newblock \bibinfo{title}{A novel method for evaluating the critical nucleus
  and the surface tension in systems with first order phase transition}.
\newblock \emph{\bibinfo{journal}{J. Chem. Phys.}}
  \textbf{\bibinfo{volume}{127}}, \bibinfo{pages}{214703}
  (\bibinfo{year}{2007}).

\bibitem{nucleation:franz05}
\bibinfo{author}{Franz, S.}
\newblock \bibinfo{title}{First steps of a nucleation theory in disordered
  systems}.
\newblock \emph{\bibinfo{journal}{J. Stat. Mech.}}
  \textbf{\bibinfo{volume}{2005}}, \bibinfo{pages}{P04001}
  (\bibinfo{year}{2005}).

\bibitem{mosaic:dzero05}
\bibinfo{author}{Dzero, M.}, \bibinfo{author}{Schmalian, J.} \&
  \bibinfo{author}{Wolynes, P.~G.}
\newblock \bibinfo{title}{Activated events in glasses: The structure of
  entropic droplets}.
\newblock \emph{\bibinfo{journal}{Phys. Rev. B}} \textbf{\bibinfo{volume}{72}},
  \bibinfo{pages}{100201} (\bibinfo{year}{2005}).

\bibitem{self:prl07}
\bibinfo{author}{Cavagna, A.}, \bibinfo{author}{Grigera, T.~S.} \&
  \bibinfo{author}{Verrocchio, P.}
\newblock \bibinfo{title}{Mosaic multistate scenario versus one-state
  description of supercooled liquids}.
\newblock \emph{\bibinfo{journal}{Phys. Rev. Lett.}}
  \textbf{\bibinfo{volume}{98}}, \bibinfo{pages}{187801}
  (\bibinfo{year}{2007}).

\bibitem{soft-spheres:bernu87}
\bibinfo{author}{Bernu, B.}, \bibinfo{author}{Hansen, J.~P.},
  \bibinfo{author}{Hiwatari, Y.} \& \bibinfo{author}{Pastore, G.}
\newblock \bibinfo{title}{Soft-sphere model for the glass transition in binary
  alloys: Pair structure and self-diffusion}.
\newblock \emph{\bibinfo{journal}{Phys. Rev. A}} \textbf{\bibinfo{volume}{36}},
  \bibinfo{pages}{4891--4903} (\bibinfo{year}{1987}).

\bibitem{soft-spheres:roux89}
\bibinfo{author}{Roux, J.-N.}, \bibinfo{author}{Barrat, J.-L.} \&
  \bibinfo{author}{Hansen, J.-P.}
\newblock \bibinfo{title}{Dynamical diagnostics for the glass transition in
  soft-sphere alloys}.
\newblock \emph{\bibinfo{journal}{J. Phys.: Condens. Matt.}}
  \textbf{\bibinfo{volume}{1}}, \bibinfo{pages}{7171--7186}
  (\bibinfo{year}{1989}).

\bibitem{cs:mezard06}
\bibinfo{author}{M\'ezard, M.} \& \bibinfo{author}{Montanari, A.}
\newblock \bibinfo{title}{Reconstruction on trees and spin glass transition}.
\newblock \emph{\bibinfo{journal}{J. Stat. Phys.}}
  \textbf{\bibinfo{volume}{124}}, \bibinfo{pages}{1317--1350}
  (\bibinfo{year}{2006}).

\bibitem{mosaic:franz07}
\bibinfo{author}{Franz, S.} \& \bibinfo{author}{Montanari, A.}
\newblock \bibinfo{title}{Analytic determination of dynamical and mosaic length
  scales in a kac glass model}.
\newblock \emph{\bibinfo{journal}{J. Phys. A: Math. Theor.}}
  \textbf{\bibinfo{volume}{40}}, \bibinfo{pages}{F251--F257}
  (\bibinfo{year}{2007}).

\bibitem{mosaic:franz07b}
\bibinfo{author}{Franz, S.}, \bibinfo{author}{Parisi, G.} \&
  \bibinfo{author}{Ricci-Tersenghi, F.}
\newblock \bibinfo{title}{Mosaic length and finite interaction-range effects in
  a one dimensional random energy model}.
\newblock \bibinfo{howpublished}{arXiv:0711.4780v2} (\bibinfo{year}{2007}).

\bibitem{glassthermo:adam65}
\bibinfo{author}{Adam, G.} \& \bibinfo{author}{Gibbs, J.~H.}
\newblock \bibinfo{title}{On the temperature dependence of cooperative
  relaxation properties in glass-forming liquids}.
\newblock \emph{\bibinfo{journal}{J. Chem. Phys.}}
  \textbf{\bibinfo{volume}{43}}, \bibinfo{pages}{139--146}
  (\bibinfo{year}{1965}).

\bibitem{spin-glass:fisher86}
\bibinfo{author}{Fisher, D.~S.} \& \bibinfo{author}{Huse, D.~A.}
\newblock \bibinfo{title}{Ordered phase of short-range ising spin-glasses}.
\newblock \emph{\bibinfo{journal}{Phys. Rev. Lett.}}
  \textbf{\bibinfo{volume}{56}}, \bibinfo{pages}{1601--1604}
  (\bibinfo{year}{1986}).

\bibitem{mosaic:stevenson06}
\bibinfo{author}{Stevenson, J.~D.}, \bibinfo{author}{Schmalian, J.} \&
  \bibinfo{author}{Wolynes, P.~G.}
\newblock \bibinfo{title}{The shapes of cooperatively rearranging regions in
  glass-forming liquids}.
\newblock \emph{\bibinfo{journal}{Nature Phys.}} \textbf{\bibinfo{volume}{2}},
  \bibinfo{pages}{268--274} (\bibinfo{year}{2006}).

\bibitem{self:pre01}
\bibinfo{author}{Grigera, T.~S.} \& \bibinfo{author}{Parisi, G.}
\newblock \bibinfo{title}{Fast monte carlo algorithm for supercooled soft
  spheres}.
\newblock \emph{\bibinfo{journal}{Phys. Rev. E}} \textbf{\bibinfo{volume}{63}},
  \bibinfo{pages}{045102} (\bibinfo{year}{2001}).

\end{thebibliography}

\end{document}